\documentclass{snmp2001}

\begin{document}

\FirstPageHeading{Bindu}

\ShortArticleName{Symmetries and Integrability Properties}

\ArticleName{Symmetries and Integrability Properties\\
of  Generalized  Fisher Type Nonlinear Diffusion \\
Equation}

\Author{P.S. BINDU and M. LAKSHMANAN}

\AuthorNameForHeading{P.S. Bindu and M. Lakshmanan}

\AuthorNameForContents{BINDU P.S. and LAKSHMANAN M.}

\Address{Centre for Nonlinear Dynamics, Department of Physics,
Bharathidasan University, \\ Tiruchirapalli - 620 024, India}
\Email{psbindu@bdu.ernet.in, lakshman@bdu.ernet.in}

\Abstract{Nonlinear reaction-diffusion systems are known to
exhibit very many novel spatiotemporal patterns.  Fisher equation
is a prototype of diffusive equations. In this contribution we
investigate the integrability properties of the generalized Fisher
type equation to obtain physically interesting solutions using Lie
symmetry analysis. In particular, we report several
 travelling wave patterns, static patterns
and localized structures depending upon the choice of the parameters involved.}

\section{Introduction}
Nonlinear partial differential equations are frequently used to model a
wide variety of phenomena in physics, chemistry, biology and other fields
\cite{bindu:whitham,bindu:scott1,bindu:murray}.
In such models, when large aggregates of microstructures consisting of
particles, atoms, molecules, defects, dislocations, etc. are able to move
and/or interact, the evolution of the concentration of the species can be shown
to obey nonlinear diffusion equations of reactive type. These equations play
an important role in dissipative dynamical systems. Many interesting physical
phenomena, such as wall propagation in liquid crystals, nerve impulse
propagation in
nerve fibres, pattern formation in dissipative systems, nucleation kinetics and
neutron action in the reactor, are closely connected with the study of nonlinear
diffusion equations. The underlying systems give rise to very many simple/complex
patterns which are essentially
distinct structures on a suitable space-time scale and they arise as  collective and
cooperative phenomena due to the underlying large number of constituent
subsystems. These structures tell us a lot about the dynamics as well as about the microscopic
behaviour of the underlying systems to some extent. As the interactions among
the constituents are nonlinear, novel structures which can mimic naturally
occurring patterns arise. These structures can be stationary or changing with
time.

\begin{figure}[th]
\begin{center}
\epsfig{file=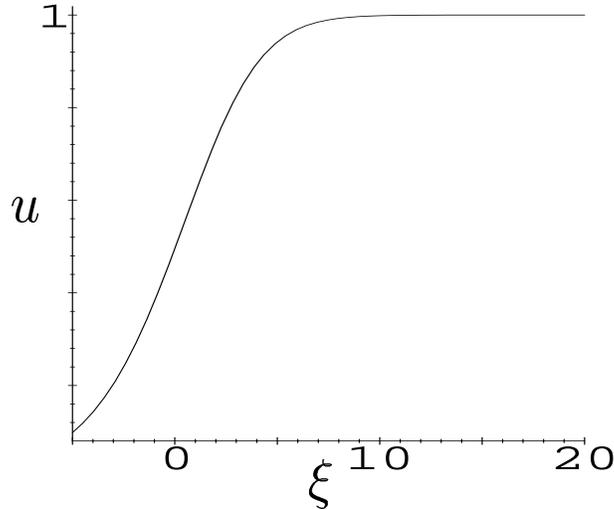, width = 0.5\linewidth}
\end{center}
\vspace{-6mm}

\caption{An exact wavefront solution~\cite{bindu:ablowitz1} of the Fisher
equation
  $\left(\xi = {x-\frac{5}{\sqrt6}t}\right)$.}
\label{bindu:front2d}
\end{figure}

 Generally, in the study of dissipative systems, one of the challenging
problems is the selection mechanism. That is, one would like to know the kinds
of evolving velocity and emerging patterns that would be selected in a kinetic
process when the system is suddenly quenched into an unstable state.
Aronson and Weinberger's work on the Fisher type nonlinear diffusion equation
\cite{bindu:aronson1} has shown the existence of distinct
selection mechanism, that is the solution $u(x,t)$ of the
Fisher equation in $(1+1)$ dimensions,
\begin{eqnarray}
\label{bindu:equation1}
u_t=u_{xx}+u(1-u),
\end{eqnarray}
converges to a local travelling wave with a definite speed from a wide class
of initial data. Further it is
known that equation (\ref{bindu:equation1}) has a travelling wave solution called
a cline \cite{bindu:murray} which is nothing
but a wave travelling in the $x$-direction with $c\ge c_{\min} =2$.
However,
the first explicit analytic form for a cline solution was obtained by Ablowitz
and Zeppetella \cite{bindu:ablowitz1}, who showed that an
 exact propagating wavefront solution
(see  Fig.~\ref{bindu:front2d}) is
of the form
\begin{gather}
\label{bindu:equation2} u(x,t)=1-\left[1+\frac{k}{\sqrt{6}}
\exp\left(\frac{x-\frac{5}{\sqrt6}t}{\sqrt 6}\right)\right]^{-2},
\end{gather}
 where $k$ is an arbitrary constant.
Here the authors made use of the
Painlev\'e  singularity structure analysis of equation~(\ref{bindu:equation1})
to find the exact solution
in the year 1979.

There is
continuing interest in recent literature \cite{bindu:brazhnik}
to investigate more general forms of the
Fisher equation. For instance,
there is an  interesting generalization of
the Fisher equation in the description of bacterial colony growth, chemical
kinetics and many other natural  phenomena and it is of the general
form \cite{bindu:grimson}
\begin{eqnarray}
\frac{\partial u(\vec{r},t)}{\partial t} = D\triangle u(\vec{r},t)
+\Lambda(u)[\nabla u(\vec{r},t)]^2+\lambda u(\vec{r},t)
G(u,\vec{r},t), \label{bindu:eqn3}
\end{eqnarray}
where $D$ is the diffusion coefficient, $\Lambda$ is the nonlocal
growth rate, $\lambda$ is the local growth rate, $G(u,\vec{r},t)$ is
the local growth function, $\nabla$ and $\triangle$ are gradient
and Laplacian operators respectively. As a special case of
equation (\ref{bindu:eqn3}), we obtain the generalized Fisher type
equation
\begin{eqnarray}
\label{bindu:equation3}
u_t - \triangle u-\frac{m}{1-u} ({\nabla u})^2 - u(1-u)=0,
\end{eqnarray}
where the subscript denotes partial differentiation with respect
to time. In the study of population dynamics, $u(\vec{r},t)$
refers to the population density at point $\vec{r}$ at time $t$.
In equation~(\ref{bindu:equation3}), the linear term modelling the
birth rate gives rise to an exponential growth in time while the
quadratic term that models competition between individuals for
food, etc. leads to a stable, homogeneous value $u=1$ at long
times and the diffusion term models the spatial variation of the
population. This introduces the possibility of spatial pattern
formation between the homogeneous regions with $u=1$ and $u=0$ for
appropriate initial conditions. Further,
 the classical Fisher equation ($m=0$) occurs in models of
population growth~\cite{bindu:murray}, neurophysiology
\cite{bindu:tuckwell},
  Brownian motion~\cite{bindu:bramson}
and nuclear reactors~\cite{bindu:canosa}. Besides allowing for
exact solutions, the $m=2$ case finds its application in real
systems such as the bacterial colony growth~\cite{bindu:grimson}
where the square-gradient term corresponds to the nonlocal growth
occuring at concentration gradients which is similar to the
nonlinear  terms in the Kuramoto--Sivashinsky equation for
propagating flame and in the theory of growing interfaces.
Moreover, models which admit exact solutions are of considerable
importance for understanding general behaviour of nonlinear
dissipative systems. In one dimensional space, such models have
received considerable attention. But many realistic models are two
or three dimensional in nature and in this direction,
 Brazhnik and Tyson \cite{bindu:brazhnik} considered
equation (\ref{bindu:equation3}) in two spatial dimensions and explored
 five kinds
of travelling wave
patterns namely plane, $V$ and $Y$-waves, a separatrix and space oscillating
propagating structures.
All these structures were found  when the medium is
unbounded and spatially homogeneous. Further they show that when the medium is
bounded and no flux is allowed through the boundaries, only plane and
oscillating waves survive because the frontline of the wave must approach
the boundary orthogonally.

In general, obtaining solutions for reaction-diffusion systems is
more complex than that for pure dispersive systems. For the latter
there are several analytical methods like the inverse scattering
transform method~\cite{bindu:novikov}, the Hirota method
\cite{bindu:hirota}, B\"acklund transformation method
\cite{bindu:ablowitz2}, Lie--B\"acklund symmetries method  and so
on.  On the other hand, for nonlinear diffusive systems, no such
formal techniques are available to solve them analytically. Very
often perturbation analysis or numerical techniques are used to
treat them. There is therefore an urgent need to isolate and
identify integrable nonlinear reaction-diffusion systems which can
act as model systems  to deal with more complicated cases. In this
connection, symmetry analysis can play a very crucial role.

Consider for example, the well known case of Burgers equation
\[
u_t=\nu u_{xx}+uu_x,
\]
where $\nu$ is the diffusion coefficient. It can be considered to
be integrable in the sense that it is linearizable: Under the
Cole--Hopf transformation $u=-\nu v_x/v$, it reduces to the linear
heat equation. It possesses interesting Lie point symmetry
structures and infinite number of Lie--B\"acklund symmetries. So,
it will be quite interesting to know about other such integrable
reaction-diffusion equations and the role of symmetries that
allows the system to exhibit different spatiotemporal patterns and
structures which usually possess some kind of symmetry. In this
direction,  the method of Lie groups is the most powerful method
to analyse nonlinear partial differential equations (PDEs) and
hence we make use of it  and the singularity structure analysis to
investigate the integrability properties and hence the
dynamics/patterns of (\ref{bindu:equation3}). We report in this
paper that the $m=2$ case of equation (\ref{bindu:equation3})
possesses infinite dimensional Lie symmetry structure, which
allows one to linearize it both in $(1+1)$ and $(2+1)$ dimensions
and to obtain a large class of exact solutions. We also obtain
several exact solutions for the $m\ne2$ case.

The plan of the paper is as follows.
In Section~2, we briefly recall some of the important reaction-diffusion
equations exhibiting
novel/complex patterns.  Then in Section~3, by carrying out the
singularity structure analysis, we point out that the PDE~(\ref{bindu:equation3})
is free from movable critical singular manifolds for the specific value $m=2$.
More interestingly, we point out that the B\"acklund
transformation deduced from the Laurent expansion gives rise
to the linearizing transformation  for this case in a natural way. In Sections~4 and~5, we
discuss different underlying patterns via symmetry analysis and similarity
reductions for the generalized Fisher type equation in 1- and 2-spatial
dimensions, respectively. Finally we summarize our results in Section~6.

\section{Reaction-diffusion systems and various patterns}
The general form of the nonlinear reaction-diffusion equation is given
by
\begin{gather*}
\frac{\partial \underline {C}}{\partial t}= \vec{\nabla}\cdot
({D}\vec{\nabla}\underline{C} ) +
\vec{F}(\underline{C}^T,\vec{r},t),\qquad
\underline{C} = (c_1,c_2,\ldots,c_n)^T, \\
{D}={\rm diag}\,(D_1,D_2,\ldots,D_n),\qquad
\vec{F}=(f_1,f_2,\ldots,f_n)^T.
\end{gather*}
Here $\underline{C}$ represents the population or concentration
densities of the species and ${D}$ and $\vec{F}$ are, in general,
nonlinear functions of $\underline{C}$ representing the
diffusivity and the reaction kinetics respectively. In such a
case, the dynamics is dominated by the onset of patterns. Inspite
of the absence of rigorous analytical tools as in the case of
soliton systems, combined local analysis and numerical
investigations on such systems have been found to exhibit a number
of important spatiotemporal patterns.

Some of the dominant patterns exhibited by
these systems are
  homogeneous or uniform steady states,
  travelling waves,
  spiral waves,
  Turing patterns (rolls, stripes, hexagons, rhombs, etc.),
  localized structures,
 spatiotemporal chaos and so on. A few of the well known models
include the following:

\subsection{The Oregonator model}
This model explains the various features of the
Belousov--Zhabotinsky reaction and was introduced by Fields,
K\"or\"os and Noyes of University of Oregon, USA in 1972. In its
simplest version it reads as~\cite{bindu:cross}
\begin{gather}
u_{1t}=D_1\nabla^2u_1+\eta^{-1}\left [u_1(1-u_1)-\frac{bu_2(u_1-a)}{(u_1+a)}
\right ],
\nonumber\\
u_{2t}=D_2\nabla^2u_2+u_1-u_2.
\label{bindu:equation4}
\end{gather}
Here $u_1$ is the concentration of the autocatalytic species ${\rm
HBrO}_2$, $u_2$ is the concentration of the transition ion
catalyst in the oxidised state ${\rm Ce}^{3+}$ and ${\rm Fe}^{3+}$
and  $\eta$, $a $ and $b$ are parameters. This model is the most
popular among the pattern forming chemical reactions. In
particular, (\ref{bindu:equation4})~exhibits `propagating pulse
solutions' that can travel through the system without attenuation.
Besides, it admits periodic wave trains, target patterns and in
two dimensions they generate spiral waves.

\subsection{Gierer--Meinhardt model}
It describes possible interaction between an activator $a$ and a
rapidly diffusing inhibitor $h$ and is of the form
\cite{bindu:koch}
\begin{gather*}
a_t=D_a\nabla^2a+\rho_a\frac{a^2}{(1+K_aa^2)}-\mu_a a+\sigma_a,\\
h_t=D_h\nabla^2h+\rho_ha^2-\mu_h h+\sigma_h,
\end {gather*}
where $D_a$ and $D_b$ are the two diffusion coeffcients, $\rho_a$ and $\rho_b$
are the removal rates and $\sigma_a$ and~$\sigma_b$ are the
 basic production terms of the activator and inhibitor
 respectively. Further~$K_a$ corresponds to the saturation
constant. This model is mainly used in the study of the development of an organism
in biological pattern formation. They are also used to model cell
differentiation, cell movement, shape changes of cells and tissues and so on.

\subsection{Brusselator model}
Among the various reaction-diffusion type model systems, this is
one of the best
studied models for the formation of chemical patterns
theoretically~\cite{bindu:walgraef}. It is based on the chemical reactions
\begin{gather*}
A \longrightarrow X,\qquad B + X \longrightarrow Y, \qquad 2X + Y
\longrightarrow 3X,\qquad X \longrightarrow E,
\end{gather*}
where the concentration of the species $A$, $B$ and $E$ are maintained
constant. Thus they form the real constant parameters of the system. The
evolution of the active species $X$ and $Y$ can be described by
\begin{gather}
X_t=A-(B+1)X+X^2Y+D_X\nabla^2X, \nonumber\\
Y_t=BX-X^2Y+D_Y\nabla^2Y,
\label{bindu:equation5}
\end{gather}
after proper rescaling. Here $D_X$ and $D_Y$ are diffusion coefficients.
This model exhibits heterogeneous patterns through Turing instability.

\subsection{Lotka--Volterra predator-prey model}
Taking into consideration the interaction of two species in which the population
of the prey is dependent on the predator and vice-versa, the model equations~\cite{bindu:okuba}
become
\begin{gather*}
S_{1t}=D_1S_{1xx}+a_1S_1-b_1S_1S_2, \\
S_{2t}=D_2S_{2xx}-a_2S_2+b_2S_1S_2.
\end{gather*}
Here $S_1$ and $S_2$ are the population densities of prey and predator. $D_1$
and $D_2$ are the diffusivities  of the two populations, respectively. The
parameters $a_1$, $a_2$ are the linear ratio of birth and death rates of the
individual species while $b_1$, $b_2$ are the nonlinear decay and growth factors
due to interaction.

\subsection{FitzHugh--Nagumo nerve conduction model}
The Hodgkin--Huxley model describes the propagation of the
electrical impulses along the axonal membrane of a nerve fibre.
FitzHugh--Nagumo nerve conduction equation~\cite{bindu:fitzhugh} is the simplest version
of the above model and is represented by the following set of equations:
\begin{gather}
V_t=V_{xx} + V -\frac{V^3}{3} - R + I(x,t),\nonumber\\
R_t=c(V+a-bR).
\label{bindu:equation6}
\end{gather}
Here the membrane potential is $V(x,t)$, $R$ corresponds to the lumped
refractory variable and $I(x,t)$ is the external injected current. The
parameters $a$ and $b$ are positive constants while~$c$ stands for the
temperature factor. The above model has been widely used to study
various phenomena in neurophysiology and cardiophysiology. This
system exhibits
travelling wave pulses~\cite{bindu:murray}.
 In particular, the two-dimensional version of~(\ref{bindu:equation6})
admits ring wave patterns as well as  spiral wave patterns
for a variety of special initial
conditions.

As mentioned in the introduction, symmetries can play a very important role in
determining the underlying dynamics of nonlinear systems. Particularly
they can help to identify integrable cases of the above type of
reactive-diffusive systems, if they exist. As an important case
study, we now investigate integrability and symmetry properties
of the generalized  Fisher type equation~(\ref{bindu:equation3}).

\section{Singularity structure analysis}
This analysis separates out the $m=2$ case for both the $(1+1)$
and $(2+1)$ dimensions as the only system for which the Fisher
equation~(\ref{bindu:equation3}) is free from
movable critical singular manifolds satisfying the Painlev\'e property~\cite{bindu:ablowitz2}.
By locally expanding the solution in the neighbourhood of the
non-characteristic singular manifold $\phi(x,t)=0$, $\phi_x,\phi_t\ne0$
in the form of the Laurent series~\cite{bindu:bindu}
\begin{gather*}
u=\sum_{j=0}^\infty u_j\phi^{j+p},
\end{gather*}
the possible values of the power of the leading order term are
found to be
\begin{align*}
 (i) \ & p = -2, \\
(ii) \  & p = \frac{1}{1-m},\qquad m \ne 1,\\
(iii) \ & p = 0.
\end{align*}
For all these leading orders, only for the value $m=2$ the solution is
free from movable critical singular manifolds since for $p=-1$
the leading order coefficient $u_0$ becomes arbitrary besides the
arbitrary singular manifold $\phi$. In all
other cases only one arbitrary function exists for $m=2$ thereby leading to special
solutions.

More interestingly, from the Laurent series expansion if we
cut off the series at
``constant'' level term, that is $j=-p$ for the leading order
$p=1/(1-m)=-1$,  $m=2$,
one can deduce the B\"acklund
transformation that gives rise to the linearizing transformation
in a natural way. Thus, defining the relation
\begin{gather}
u=\frac{u_0}{\phi}+u_1,
\label{bindu:eqn7}
\end{gather}
we demand  that if $u_1$ is  a solution
of equation~(\ref{bindu:equation3}) for the case $m=2$, then $u$
is also a solution, from which the
B\"acklund transformation is deduced.
Now starting from the
trivial solution, $u_1=0$ of (\ref{bindu:equation3}),  we find  that the
equations
for $u_0$ and $\phi$ in equation (\ref{bindu:eqn7}) are consistent for the choice
$u_0=\phi$, giving rise to the new solution $u=1$. This is nothing
but an exact solution of equation~(\ref{bindu:equation3}). Then with
$u_1=1$ as
the new seed
solution, one can check from equations satisfied by $u_0$ and $\phi$
that
\begin{gather}
u_0=-1,\qquad \phi_t-\phi_{xx}-\phi+1=0.
\label{bindu:eqn8}
\end{gather}
Choosing $\phi = 1+\chi$, equation~(\ref{bindu:eqn8}) can be rewritten as
the linear
heat equation,
\begin{gather}
\chi_t-\chi_{xx}-\chi=0.
\label{bindu:eqn9}
\end{gather}
Thus the transformation
\begin{gather}
u=1-\frac{1}{1+\chi},
\label{bindu:eqn10}
\end{gather}
where $\chi$ satisfies the linear heat equation (\ref{bindu:eqn9}), is
the
linearizing transformation for equation~(\ref{bindu:equation1}) in $(1+1)$ dimensions for
the choice $m=2$ in an automatic way. We note that this is exactly the
transformation given in ref.~\cite{bindu:wang} in an adhoc way. Here
 we have given an
 interpretation for the transformation in terms of the B\"acklund
transformation.  The same transformation~(\ref{bindu:eqn10})
linearizes equation~(3) in $(2+1)$ dimensions (for $m=2$) as well, where $\chi$
satisfies the two
dimensional linear heat equation $\chi_t-\chi_{xx}-\chi_{yy}-\chi=0$. Further
equation~(\ref{bindu:eqn10}) transforms equation~(\ref{bindu:equation3}) in $(3+1)$
dimensions to the 3-dimensional heat equation as well; however,
we do not study the case further here.

\section{Symmetries and integrability properties\\
 of $\pbf{(1+1)}$ dimensional generalized Fisher equation}
The generalized Fisher equation (\ref{bindu:equation3}) in its
$(1+1)$ dimensional form reads as
\begin{gather}
u_t-u_{xx}-\frac{m}{1-u}-u+u^2=0.
\label{bindu:eqn11}
\end{gather}
An invariance analysis of equation (\ref{bindu:eqn11}) under the infinitesimal
transformations
\begin{gather*}
x \longrightarrow X = x+\varepsilon\xi(t,x,u), \qquad
t \longrightarrow T = t+\varepsilon\tau(t,x,u),\\
u \longrightarrow U = u+\varepsilon\phi(t,x,u),  \qquad \varepsilon \ll 1,
\end{gather*}
separates out the $m=2$ case in that it possesses a nontrivial infinite-dimensional
Lie algebra of symmetries
\begin{gather*}
 \tau  =  a, \qquad \xi  =   b, \qquad \phi  = c(t,x)(1-u)^2.
\end{gather*}
Here $ a$, $b $ are arbitrary constants and  $ c(t,x) $
is any solution of the linear heat equation $c_t-c_{xx}-c =0$. For all other
values of $m$  in equation~(\ref{bindu:eqn11}) one gets trivial translation symmetries
\begin{gather*}
\tau = a,\qquad \xi = b,\qquad \phi = 0.
\end{gather*}

In order to obtain solutions of physical importance and corresponding patterns,
we make use of the method of similarity reductions. This leads to the
similarity reduced variables for the $m=2$ case as
\begin{gather}
z=ax-bt,\qquad  u=1- \frac {a}{a+v(z)+\int c(t,x)dt}.
\label{bindu:eqn12}
\end{gather}
Using (\ref{bindu:eqn12}), equation (\ref{bindu:eqn11}) can be reduced
for the $m=2$ case to  the similarity reduced ordinary differential
equation (ODE)
\begin{gather*}
a^2 v''+bv' + v=0
\end{gather*}
whose general solution is
\begin{gather*}
v = I_1 e^{m_1z}+ I_2 e^{m_2z}, \qquad
m_{1,2}=\frac{-b \pm \sqrt {b^2-4a^2}}{2a^2},
\end{gather*}
where $I_1$ and $I_2$ are integration constants thereby leading to
\begin{gather*}
u= \left \{\!\!
    \begin{array}{lll}
\displaystyle 1-\frac {a}
{a+I_1 e^{m_1(ax-bt)}+I_2 e^{m_2(ax-bt)}+
\int c(t,x) dt}, \quad b^2-4a^2 > 0;\vspace{2mm}\\
\displaystyle 1-\frac{a}{ a+e^{p(ax-bt)} \left ( I_1+I_2(ax-bt) \right )
+\int c(t,x)dt },\quad b^2-4a^2 = 0; \vspace{2mm}\\
\displaystyle 1-\frac{a}{a+e^{p(ax-bt)} \left ( I_1 \cos {q(ax-bt)}
+ I_2 \sin {q(ax-bt)} \right )
+\int c(t,x)dt}, \quad  b^2-4a^2 < 0 \!
\end{array}
\right.
\end{gather*}
with $p = -b/2a^2$, $q = \sqrt{4a^2-b^2}/2a^2$, as the solution to the
original PDE~(\ref{bindu:eqn11}). Here the similarity reduced
variable~(\ref{bindu:eqn12}) is nothing but the
linearizing transformation~(\ref{bindu:eqn10}).

\begin{figure}[th]
\centering
\epsfig{file=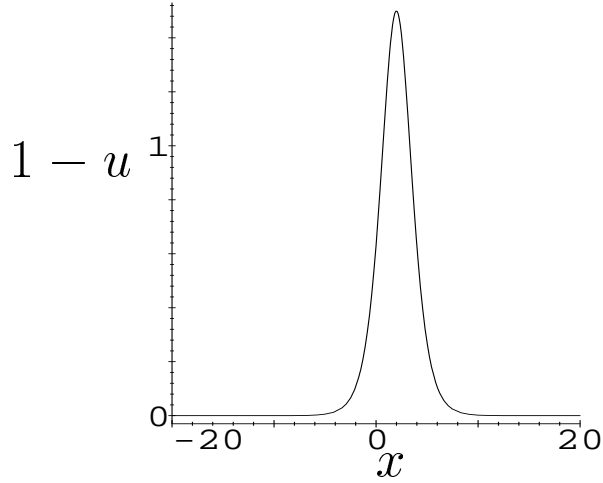, width=.5\linewidth}
\vspace{-3mm}

\caption{A static solitary wave pulse for $m=1/2$ of the generalized Fisher
equation~(\ref{bindu:eqn11}).}
\label{lakshmanan:blimit}
\end{figure}

\begin{figure}[th]
\centering
\epsfig{file=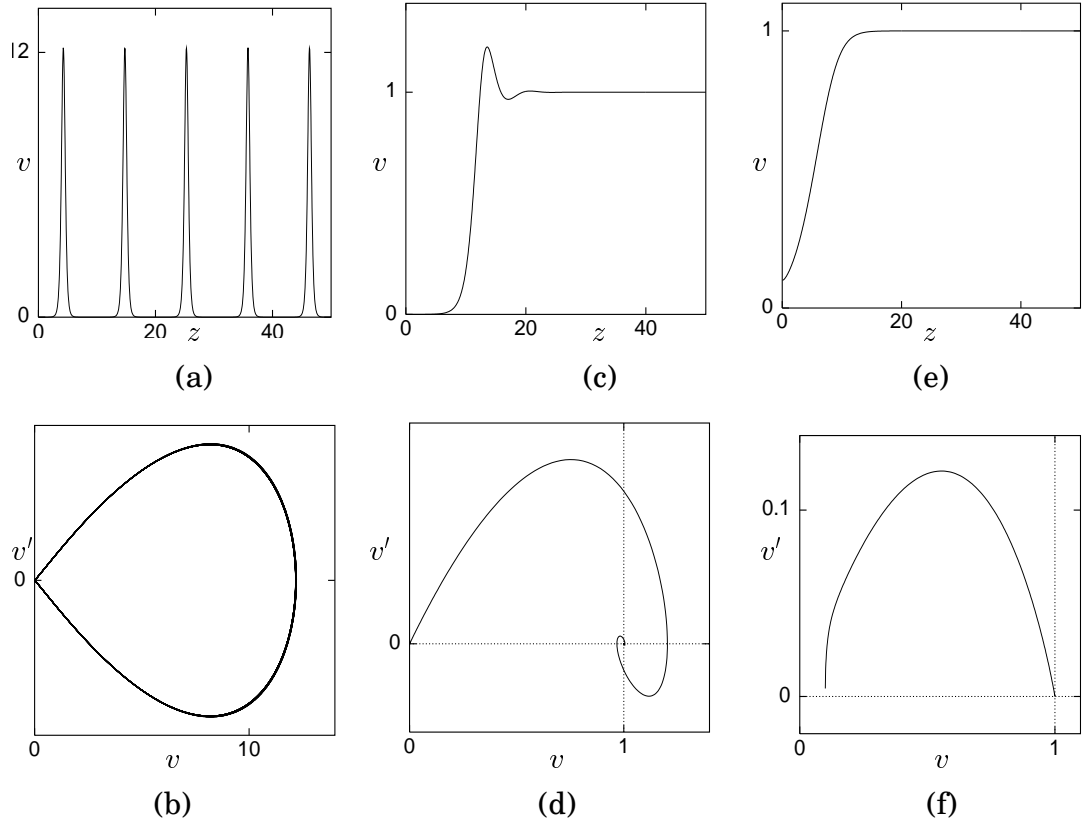, width=0.9\linewidth}
\vspace{-3mm}

\caption{Propagating patterns and corresponding phase portraits in the
$v-v'$ plane of equation~(\ref{bindu:equation13}): (a)~periodic pulses;
(b) limit cycle; (c)~travelling pulse; (d)~stable spiral;
(e)~travelling wavefront; (f)~stable node.}
\label{lakshmanan:phase}
\end{figure}

Proceeding in a similar fashion for all the other (nonintegrable)
cases ($m\ne2)$, the
similarity  variables $z=ax-bt$ and $u=w(z)$ reduce
equation~(\ref{bindu:eqn11}) to the ODE
\begin{gather}
a^2 v v'' - ma^2 v'{}^{2} + b v v' - (1- v) v^2=0, \qquad v = 1-w,
\label{bindu:equation13}
\end{gather}
which is in general nonintegrable except for  $m=0$ and $b/a=5/\sqrt{6}$.
This special choice leads to the cline solution (\ref{bindu:equation2}) obtained by Ablowitz
and Zeppetella \cite{bindu:ablowitz1}. In the static case ($b=0$),
one obtains elliptic function
solutions.
Besides, a particular solitary wave solution
\begin{gather*}
u= 1-\frac{(3-2m)}{(2-2m)}\left [ \mbox{sech} ^2 \left (I_2-\frac{x}{2}
\sqrt{\frac{1}{1-m}} \right ) \right ], \qquad m < 1
\end{gather*}
with $ I_2 $ as the second integration constant, which is a limiting
case of a
elliptic function solution, is also obtained (refer Fig.~\ref{lakshmanan:blimit}).
In the general case, as equation~(\ref{bindu:equation13}) is of nonintegrable
nature, we make use of
numerical techniques to study the underlying dynamics.
Here we obtain typical
periodic wave trains   for $b/a=0$ which is in accordance with the
fact that reaction-diffusion systems exhibiting limit cycle motion in the absence
of diffusion exhibits travelling wave patterns (Fig.~\ref{lakshmanan:phase}a,b).
For  $b/a=1$,
we get a propagating
pulse (Fig.~\ref{lakshmanan:phase}c) and the corresponding phase portrait ($v-v'$) shows a  stable
spiral equilibrium point
(Fig.~\ref{lakshmanan:phase}d). On increasing the value of $b/a$, that is, at $b/a\ge 2$
($b= 2.041$)~\cite{bindu:ablowitz1}, the system supports a
travelling wave front (Fig.~\ref{lakshmanan:phase}e) and the trajectories  in the phase plane ($v-v'$)
correspond to  a
stable node (Fig.~\ref{lakshmanan:phase}f).

\section{The $\pbf{(2+1)}$ dimensional generalized Fisher equation}
Extending a similar analysis to the $(2+1)$ dimensional case of the
generalized Fisher equation
\begin{gather}
u_t-u_{xx}-u_{yy}-\frac{m}{1-u}\left(u_x^2+u_y^2\right)-u+u^2=0,
\label{bindu:eqn13}
\end{gather}
one finds that the invariance analysis of equation~(\ref{bindu:eqn13}) under the infinitesimal
transformation singles out the special value $m=2$ for which the
Lie point symmetries are
\begin{gather*}
\tau=a,
\qquad \xi=b_3y+b_4, \qquad \eta = -b_3x+d_4,\qquad \phi = c(t,x,y)(1-u)^2,
\end{gather*}
where $\eta$ is the infinitesimal symmetry associated with the variable $y$,
$ c(t,x,y) $ is the solution of the two
dimensional linear heat equation $c_t-c_{xx}-c_{yy}-c = 0$ and
 $b_3,\;b_4$ and $d_4$ are arbitrary constants.
But for all other choices of $m$ $(\ne 2)$ we get
\begin{gather*}
\tau=a, \qquad
\xi=b_3y+b_4, \qquad \eta = -b_3x+d_4, \qquad \phi = 0.
\end{gather*}

In a similar fashion as that for the $(1+1)$ dimensional case,
the similatiry
variables for the $m=2$ case
\begin{gather}
z_1=\frac{b_3}{2}(x^2+y^2)+b_4y-d_4x,\qquad z_2=-t
-\frac{a}{b_3}\sin^{-1}\left(\frac{d_4-b_3x}{\sqrt{d_4^2+2b_3z_1+b_4^2}}
\right),\nonumber\\
  u=1-  \frac {a}{w(z_1,z_2)+\int c(t,x,y)dt}
\label{bindu:eqn14}
\end{gather}
reduce the PDE~(\ref{bindu:eqn13}) to
\begin{gather}
w_{z_2}+2b_3w_{z_1}+\left(2b_3z_1+b_4^2+d_4^2\right)w_{z_1z_1}+\frac {a^2w_
{z_2z_2}}{2b_3z_1+b_4^2+d_4^2}+w-a=0.
\label{bindu:eqn15}
\end{gather}
Here too one can obtain the linear heat equation
\begin{gather*}
\chi_t-\chi_{xx}-\chi_{yy}-\chi=0,\\
 \chi=\frac{1}{a}\left
[w(z_1,z_2)+\int c(t,x,y)dt\right ],
\end{gather*}
from the similarity form  (\ref{bindu:eqn14}). Such a transformation can be
interpreted as the linearizing transformation  from a group
theoretical point of view.

\begin{figure}[th]
\begin{center}
\epsfig{file=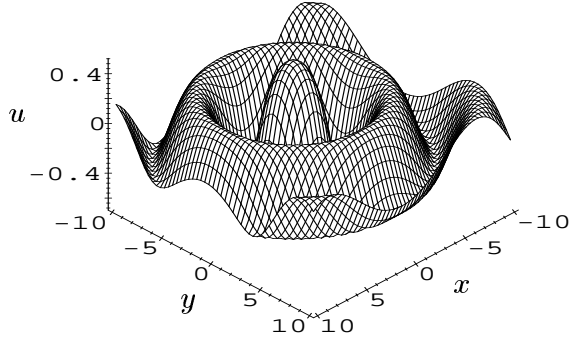, width = 0.5\linewidth}
\end{center}
\vspace{-5mm}
\caption{Circularly symmetric patterns of the $(2+1)$ dimensional generalized
Fisher equation (\ref{bindu:eqn13}) for $m=2$.}
\label{lakshmanan:bbessel}
\end{figure}

Carrying out a Lie symmetry analysis for
equation~(\ref{bindu:eqn15}) also, one can obtain the new
similarity variables
\begin{gather*}
\zeta=\bar {z}_1,
\quad {w}= a+e^{ \left (\frac {c_1\bar {z}_2}{c_3}\right
)}\left [f(\zeta)+\frac{1}{c_3} \int \hat{c}_2(\bar {z}_1,
\bar {z}_2) e^{\left(-\frac {c_1}{c_3} \bar {z}_2 \right )} d\bar {z}_2
\right ],\\
\bar {z_1} = 2b_3z_1+b_4^2+d_4^2, \qquad
\bar {z}_2 = z_2, \qquad b_3,d_4\ne 0,
\end{gather*}
where $f$ satisfies the linear second order ODE of the form
\begin{gather}
\zeta^2 f''+\zeta f'+(A+B \zeta)f=0,\nonumber\\
 A=(ac_1/2b_3c_3)^2, \qquad B = (1+c_1/c_3)/4b_3^2,
\end{gather}
with prime  denoting differentiation w.r.t.\ $\zeta$. Thus the
solution to the original PDE reads as
\begin{gather}
u=1- a \Bigg [ a + e^{\left(\frac{c_1}{c_3} \bar{z}_2\right )}
 \Bigg (I_1Z_1\left(2\sqrt{B\bar{z}_1}\right)+
I_2Z_2\left(2\sqrt{B\bar{z}_1}\right)\nonumber\\
\qquad {}-\int \frac{\hat{c}_2(\bar{z}_1,\bar{z}_2)}{c_3}
e^{\left(\frac{c_1}{c_3}\bar{z}_2\right)}
d\bar{z}_2\Bigg) +\int c(t,x,y)dt \Bigg]^{-1}.
\label{bindu:eqn16}
\end{gather}
In the limit $b_4=d_4=c_1=0$ the system is found to exhibit circularly symmetric
structures given in Fig.~\ref{lakshmanan:bbessel}.

More interestingly, in the special case $b_3=0$, $d_4=0$ the system
exhibits propagating wave structures and the corresponding forms are
\begin{gather}
u=\left \{
    \begin{array}{l}
\displaystyle 1-a\Bigg \{a+\exp\Bigg [-k\left (\frac{a}{b_4}x-t\right )\Bigg ]\Bigg [ I_1
\cos(\sqrt{k_1}c_5b_4y)+I_2\sin\left(\sqrt{k_1}c_5b_4y\right) \\
\displaystyle\mskip 150 mu{}+\int \frac{\hat{c}_3(z_1,z_2)}{c_5}e^{kz_2}dz_2 \Bigg ]
+\int c(t,x,y)dt\Bigg \}^{-1}, \quad k_1<0, \vspace{2mm}\\
\displaystyle 1-a\Bigg \{a+\exp\Bigg [ -k\left (\frac{a}{b_4}x-t \right )
\Bigg ]
\Bigg [ I_1e^{\sqrt{k_1}c_5b_4y}+I_2e^{-\sqrt{k_1}c_5b_4y} \\
\displaystyle \mskip 150 mu {}+\int \frac{\hat{c}_3(z_1,z_2)}{c_5}e^{kz_2}dz_2 \Bigg ]
+\int c(t,x,y)dt\Bigg \}^{-1}, \quad k_1>0,\vspace{2mm}\\
\displaystyle 1-a\Bigg \{a+\exp\Bigg[-k\left (\frac{a}{b_4}x-t \right )\Bigg ]
\Bigg [ I_1c_5b_4y+I_2 \\
 \displaystyle\mskip 150 mu
{}+\int \frac{\hat{c}_3(z_1,z_2)}{c_5}e^{kz_2}dz_2\Bigg ]
+\int c(t,x,y)dt \Bigg \}^{-1}, \quad k_1=0,
\end{array}
\right.
\label{bindu:eqn17}
\end{gather}
where the parameter $k_1=\frac{1}{b_4^2c_5^2}\left[k-\left(\frac{ak}{b_4}\right)^2-1\right]$
with  $k=-c_2/c_5$ and $z_1=b_4y$,
$z_2 = \frac{a}{b_4}x-t$. Here $c_2$, $c_4$, $c_5$ are arbitrary constants of
integration.
Equation~(\ref{bindu:eqn17}), in particular exhibits the five classes of
bounded travelling wave solutions reported by Brazhnik and
 Tyson~\cite{bindu:brazhnik} for certain choice
of the parameters involved along with the specific assumptions of the
functions $\hat{c}_3(z_1,z_2) = 0$ and $c(t,x,y) = 0$.
The corresponding  solutions are
given below.

\begin{figure}[th]
\begin{center}
\epsfig{file=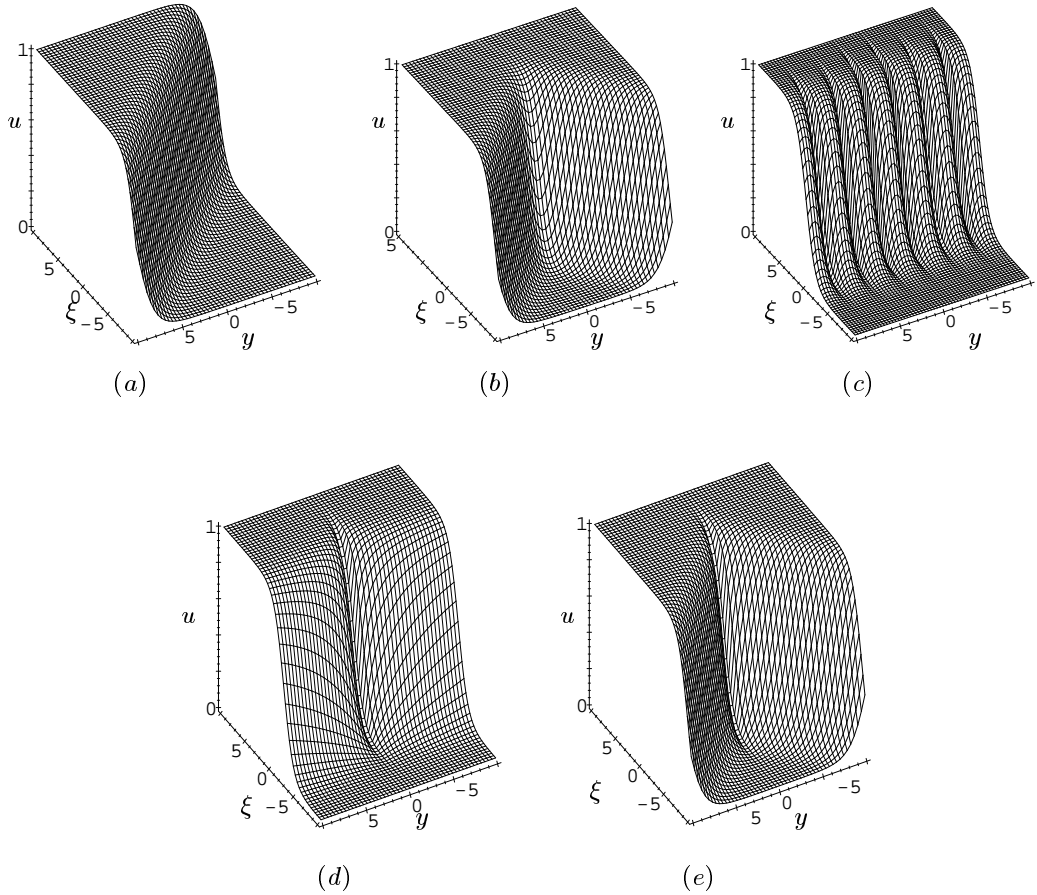,width=0.9\linewidth}
\end{center}

\vspace{-6mm}

\caption{Five interesting classes of propagating wave patterns as obtained
in ref.~\cite{bindu:brazhnik}, which follow from equations (\ref{bindu:eqn17}): (a)~travelling
waves; (b)~V-waves; (c)~oscillating front; (d)~separatrix solution;
(e)~Y-waves with $\xi=-k(\frac{a}{b_4}x-t)$.}
\label{lakshmanan:waves}

\end{figure}

Among the classes of solutions, the simplest travelling wave
solution (Fig.~\ref{lakshmanan:waves}a)
\begin{gather*}
u=1-\frac{1}{1 + A \exp \left [-k\left (\frac{a}{b_4}x-t\right )\pm\sqrt{k_1}c_5b_4y \right  ]},
\qquad k_1>0
\end{gather*}
can be constructed by assuming either $I_1 = 0$ or $ I_2 = 0$.
For $I_1=I_2$ $(\ne0)$, we obtain a V-wave pattern (Fig.~\ref{lakshmanan:waves}b)
\begin{gather*}
u=1-\frac{1}{1 + A \exp \left [-k\left
(\frac{a}{b_4}x-t\right )\right ]\cosh\left(\sqrt{k_1}c_5b_4y\right)},
\qquad k_1>0.
\end{gather*}
Again the case $I_2=0$ and $k_1 < 0$
leads to a wave front oscillating in space  (Fig.~\ref{lakshmanan:waves}c) and is represented by
\begin{gather*}
u=1-\frac{1}{1 + A \exp \left[-k\left
(\frac{a}{b_4}x-t\right )\right ]\left|\cos\left(\sqrt{k_1}c_5b_4y\right)\right|}.
\end{gather*}
But
when $I_1\ne 0$ and $I_2=0$ we get a separatrix (Fig.~\ref{lakshmanan:waves}d)
\begin{gather*}
u=1-\frac{1}{1 + A |y|\exp \left[-k\left (\frac{a}{b_4}x-t\right )\right ]}.
\end{gather*}
 Finally  for positive $k_1$ and $ I_1=-I_2 $ the Y-wave solution
(Fig.~\ref{lakshmanan:waves}e)
becomes
\begin{gather*}
u=1-\frac{1}{1 + A \exp\left[-k\left(\frac{a}{b_4}x-t\right)\right ]
\left|\sinh\left(\sqrt{k_1}c_5b_4y\right)\right|}.
\end{gather*}

In each of  the above solutions $A$ is a positive constant.
It is a well known fact about  Fisher equation is that it forms a basis for
many nonlinear models of different
nature. As a result, the above solutions are reminiscent of patterns
from different
fields. In particular, V-waves are characterized in the framework of
geometrical crystal growth related
models~\cite{bindu:schwendeman} and in excitable media~\cite{bindu:brazhnik1}
while  space oscillating fronts are relevant to cellular flame structures
and patterns in chemical reaction diffusion systems~\cite{bindu:scott}.
Further it has been shown in~\cite{bindu:brazhnik2} with a geometrical
model that excitable media
can support space-oscillating fronts.
Several static structures can also be obtained as
limiting cases of the above solutions~(\ref{bindu:eqn16}) and~(\ref{bindu:eqn17}).

Finally, a similar analysis for the nonintegrable~ ($m\ne2$) case
yields static patterns/structures in $(x,y)$ variables. Here one
has to look for certain special solutions due to its nonintegrable
nature. That is, for $b_3=0$ and $d_4=0$ with the similarity
variables  $z_1=b_4y$, $z_2= \frac{a}{b_4}x-t$, $u =w(z_1,z_2)$,
the reduced ODE reads as
\begin{gather*}
Df''+\frac{Dm}{1-f}{f'}^2-c_1f'+f(1-f)=0, \\
D=\left (\frac{a^2}{b_4^2}{c_1}^2+b_4^2{c_2}^2\right ),
\qquad ()'=d/d\zeta,
\end{gather*}
with $\zeta=-c_1\left(\frac{a}{b_4}x-t\right)+c_2b_4y$ and $w = f(\zeta)$,
giving rise to plane wave structures.
For $b_3=0$, the similarity variables $ z_1=d_4x-b_4y$,
$z_2=ax-b_4t$ and $u=w(z_1,z_2)$
 reduces the PDE to an ODE
\begin{gather*}
Af_1''+Bf_1'-\frac{Am}{f_1}f_1^{'2}-f_1+f_1^2=0, \qquad ()'=d/d\zeta
\end{gather*}
with $ f_1=1-f$, $A=a^2\left(c_1^2d_4^2+c_2^2b_4^2\right)$,
$B=-d_4b_4(c_1+c_2)$ and $\zeta = ac_2(d_4x-b_4y)-d_4(c_1+c_2)
(ax-b_4t)$, $w=f(\zeta)$. Then the  system is found to possess
elliptic function solutions
including the limiting case of the solitary pulse
for certain choices of the constants involved.

\section{Conclusion}
Our studies on the integrability/symmetry properties of the
 the generalized Fisher type nonlinear reaction-diffusion
equation show that the system under consideration possesses interesting Lie
point symmetries that could form infinite dimensional Lie algebra for the
particular choice of the system parameter $m=2$, thereby exhibiting
various interesting patterns and dynamics. Besides, the singularity structure
analysis singles out the
$m=2$ case as the only system parameter for which the generalized Fisher type
equation is free from movable critical singular manifolds. The
generalized Fisher equation is found to possess a large number of interesting
wave patterns. It will be of interest to consider other physically interesting
reaction-diffusion systems from the Lie symmetry point of view and to study the
underlying patterns.

\subsection*{Acknowledgements}
This work forms a part of the National Board of Higher Mathematics,
Department of Atomic Energy, Government of India and the Department
of
Science and Technology, Government of India research projects.

\LastPageEnding

\end{document}